\documentstyle[seceq,epsbox,preprint]{jpsj}

\def\runtitle
{Numerical Study of Aging in the Generalized Random Energy Model}
\def\runauthor{Munetaka {\sc Sasaki} and Koji {\sc Nemoto}}
\title{Numerical Study of Aging in the Generalized Random Energy Model}
\author{Munetaka {\sc Sasaki} and Koji {\sc Nemoto}}

\inst{Division of Physics, Hokkaido University, Sapporo 060-0810}
\recdate{\phantom{February 28, 2000}}
\abst{
Magnetizations are introduced to the Generalized Random Energy Model (GREM) 
and numerical simulations on ac susceptibility is made for 
direct comparison with experiments in glassy materials. 
Prominent dynamical natures of spin glasses, {\it i.e.}, 
{\em memory} effect and {\em reinitialization}, 
are reproduced well in the GREM. The existence of many layers causing 
continuous transitions is very important for the two natures. 
Results of experiments in other glassy materials such as 
polymers, supercooled glycerol and orientational glasses, 
which are contrast to those in spin glasses, are interpreted well 
by the Single-layer Random Energy Model. 
}

\kword{aging, generalized random energy model, memory effect, 
reinitialization, temperature specific dynamics, cumulative dynamics
}
\begin{document}
\sloppy
\maketitle
\section{Introduction}\label{sec:introduction}
In spin glasses, it is well known that dynamical behavior strongly depends on 
history of the system after quenching from above the transition temperature 
$T_{\rm c}$. These phenomena are called aging and have been studied with 
various experimental protocols such as 
the isothermal\cite{isothermal1,isothermal2}, 
the \( T \)-shift\cite{TS1,TS2} and the \( T \)-cycle 
one\cite{TC1,TC2,TC3,TC4,TC5}. From the theoretical point of view, 
aging phenomena have been studied 
along two different pictures, {\it i.e.}, so-called droplet 
picture\cite{droplet1,droplet2,droplet3,droplet4} 
and hierarchical picture\cite{TC1,TC3,TC4}. 
The Generalized Random Energy Model (GREM)\cite{GREM1,GREM2,GREM3,GREM4} 
is a model that belongs to the latter. 
This model has a hierarchical structure causing 
continuous transitions as the system is cooled down. 
These continuous transitions correspond to 
successive branching process of free energy in the 
hierarchical picture. Bouchaud and Dean \cite{GREM1} have shown that 
aging naturally occurs in this model and 
the time correlation function \( C(t+t_{\rm w},t_{\rm w}) \) 
satisfies a \( t/t_{\rm w} \) scaling law. 
Furthermore, we have recently made simulations\cite{condmat,condmat2} 
similar to experiments on aging phenomena with temperature variations. 
As the consequence, it has been 
shown that results of experiments are reproduced well 
according to the hierarchical picture.

Recently, Jonason {\it et al}\cite{Jonason} 
have made a new experiment, in which curious dynamical natures of 
spin glasses are observed very clearly. 
This experiment consists of the following two runs. 
In the first run, the sample is continuously cooled from 
\( T_{\rm max} \) ($>T_{\rm c}$) to \(  T_{\rm min} \) ($<T_{\rm c}$) 
at a constant rate, and is immediately reheated at the same rate. 
During the cooling and the reheating, out-of-phase 
ac-susceptibility \( \chi'' \) is measured as a function of 
temperature. We call this curve as \( \chi_{\rm ref}'' \). 
The difference between the observed value in the cooling and that 
in the reheating is not observed in spin glasses, 
while the hysteresis is observed in other glassy materials, 
such as polymer glasses\cite{PG}, orientational glasses\cite{OG} and 
disordered ferromagnets\cite{DF}. In the second run, 
the sample is cooled from \( T_{\rm max} \) to a waiting temperature 
\( T_{\rm wait} \) (\( T_{\rm min} < T_{\rm wait}< T_{\rm c}\)), 
and is kept at \( T_{\rm wait} \) during a certain time interval. 
The sample ages and \( \chi '' \) decreases during the interval. 
Then the system is cooled to \( T_{\rm min} \) and is reheated to 
\( T_{\rm max} \) without any stops. 
Hereafter this curve is denoted as \( \chi_{\rm wait}'' \). 

An important result of this experiment is that 
\( \chi_{\rm wait}'' \) merges with \( \chi_{\rm ref}'' \) at 
very beginning of the resumed cooling as if the system forgets the aging 
at \( T_{\rm wait} \) ({\em reinitialization} of aging). 
But this aging is still imprinted and a dip of  \( \chi_{\rm wait}'' \) 
created in the cooling stage is exactly recovered in the reheating stage 
({\em memory} effect). As Hammann {\it et al} have 
pointed out\cite{comparative}, these two effects make dynamics 
in spin glasses {\em temperature specific} in the sense that the effect of 
aging at \( T_{\rm wait} \) (or the difference between $\chi_{\rm ref}''$ 
and $\chi_{\rm wait}''$) only appears near \( T_{\rm wait} \) in the 
both cooling and reheating stages. The purpose of this manuscript is 
to study aging phenomena of the GREM with this protocol. 

The organization of this manuscript is as follows: 
In \S 2, the GREM is explained and magnetizations are introduced to this model.
In \S 3, the results of the simulations are presented. 
In \S 4, it is discussed how a variety of aging phenomena 
observed in glassy materials are interpreted within the GREM.

\section{Model}\label{sec:Model}
The GREM is schematically shown in Fig.~1. 
This model consists of $L$ layers which are piled up hierarchically. 
The bottom points represent accessible states of the system and each branch 
represents a barrier over which the system goes to another state. 
Each branching point has \( N \) branches, so that the number of 
states is \( N^L \). It is assumed that $N$ is large enough. 
Energy barriers of the \( n \)-th layer counted from the bottom, \( E_n \), 
are given randomly and independently according to the distribution
\begin{equation}
\rho_n(E_n)=\frac{1}{T_{\rm c}(n)}\exp\left[-\frac{E_n}{T_{\rm c}(n)}\right],
\end{equation}
where \( T_{\rm c}(n) \) is the transition temperature of the 
\( n \)-th layer. From the distribution, the averaged relaxation time 
of the $n$-th layer \( \langle \tau(n) \rangle \) is easily calculated as 
\begin{eqnarray} 
\langle \tau(n) \rangle &\equiv& \int_0^{\infty} {\rm d}E_n \rho(E_n)\tau_0 
\exp(E_n/T)\nonumber \\
&=&\left\{ 
  \begin{array}{cl}
\displaystyle{\frac{T\tau_0}{T-T_{\rm c}(n)}} &\mbox{($ T>T_{\rm c}(n) $)}, \\
\infty&\mbox{($ T\le T_{\rm c}(n) $)},
\end{array}\right.
\end{eqnarray}
where \( \tau_0 \) is a microscopic time scale. This means that a transition 
from the ergodic phase to the non-ergodic phase occurs at \(T_{\rm c}(n)\) 
in the \( n \)-th layer. The transition temperatures are chosen so as to 
satisfy the inequality \( T_{\rm c}(1)<T_{\rm c}(2)<\cdots <T_{\rm c}(L) \). 
Therefore, transitions occur continuously 
from the uppermost (the \( L \)-th) layer to the lowest one.

Next, let us explain how magnetizations are introduced to the 
GREM\cite{condmat,condmat2}. 
It is natural to assume that the nearer two states 
$\alpha$ and $\beta$ are the stronger the correlation between 
the two magnetizations is, and that the distance between the two states 
\( d(\alpha,\beta) \) is measured by the layer from which they 
are separated, {\it e.g.}, in Fig.~1 $d(\alpha,\beta)=1$ and 
$d(\alpha,\gamma)=2$. To incorporate these aspects, 
we assign the value of the magnetization $M_\alpha$ to state $\alpha$ as
\begin{equation}
M_{\alpha} ={\cal M}_0(\alpha)+{\cal M}_1(\alpha_1)+\cdots
+{\cal M}_{L-1}(\alpha_{L-1}),
\end{equation}
where $\alpha_k$ is the $k$-th ancestor of $\alpha$ and 
\( {\cal M}_k(\alpha_k) \) is a contribution 
from the branching point. The value of \( {\cal M}_k(\alpha_k) \) is 
given independently and randomly from distribution 
$D_k({\cal M}_k)$ with the mean value \( \overline{{\cal M}_k}=0 \). 
If $d(\alpha,\beta)=k$, 
the correlation between $M_\alpha$ and $M_\beta$ comes from the 
common contributions of ${\cal M}_n$ $(n=k,k+1,\cdots)$ to these magnetizations
\begin{equation}
\overline{M_{\alpha}M_{\beta}}=\sum_{n=k}^{L-1}
\overline{{\cal M}_n^2}.
\end{equation}
It decreases monotonically as $k$ increases and the barrier between two states 
becomes higher, which is observed in the SK model\cite{Nemoto}. 
In the present simulation, the same uniform distribution is chosen 
for ${\cal M}$ of all the layers as
\begin{equation} 
D_n({\cal M}_n) = \left\{ 
  \begin{array}{cl}
\frac{\sqrt{L}}{2} &\mbox{($|{\cal M}_n| \leq \frac{1}{\sqrt{L}} $)}, \\
0 &\mbox{($ |{\cal M}_n| > \frac{1}{\sqrt{L}} $)}.
\end{array}\right.
\end{equation}
The range of the distribution is chosen so that 
the variance of $M_\alpha$ is independent of $L$.

The dynamics of the system is described by the master equation for the 
probability \( P_{\alpha}(t) \) of finding the system at a state \( \alpha \) 
at time \( t \),
\begin{equation}
\frac{\rm d}{{\rm d}t} P_{\alpha}(t)
=\sum_{\beta\ne\alpha}W_{\alpha \beta}(t)P_{\beta}(t)-
\sum_{\beta\ne\alpha}W_{\beta \alpha}(t)P_{\alpha}(t),
\end{equation}
where $W_{\alpha \beta}(t)$ is the transition rate at \( t \)
for going from \( \beta \) to \( \alpha \) in a unit time. 
The uniform distribution is chosen for the initial condition, {\it i.e.}, 
\( P_{\alpha}(0)=N^{-L} \) for each \( \alpha \). 
This means that the system is quenched from an infinitely high temperature. 
The transition rate $W_{\alpha \beta}(t)$ is given as
\begin{equation}
W_{\alpha \beta}(t)=\tau_0^{-1}\sum_{k=d(\alpha,\beta)}^L 
\frac{1}{N^k}
\left\{ \exp\left[-\frac{\sum_{n=1}^k E_{n}(\beta)+H(t)M_{\beta}}
{T(t)}
\right]
-\exp\left[-\frac{\sum_{n=1}^{k+1} E_{n}(\beta)+H(t)M_{\beta}}{T(t)}
\right] \right\},
\end{equation}
where $H(t)$ is magnetic field and \( E_{L+1} \) is hypothetical 
energy whose value is infinity. 
The factor in the braces on the right hand represents the 
probability that the system can be activated to the $k$-th layer but not to 
the $k+1$-th layer, and the factor $1/N^k$ represents the probability that 
the system falls into $\alpha$. The details of how the dynamics 
is simulated in the limit $N\rightarrow\infty$ are presented in 
ref.~\citen{condmat}. 

\section{Results of Simulations}
The simulation is done in the GREM with $L=2$, 
$T_{\rm c}(1)=0.5$ and $T_{\rm c}(2)=1.0$. 
The amplitude and the period of the applying ac-field are 
\( 0.1 \) and $100 \tau_0$, respectively. 
The system is cooled 
at the rate of $2.0\times10^{-5} T/\tau_0$, and is immediately reheated 
at the same rate in the case of \( \chi_{\rm ref}'' \) measurement. 
The cooling is intermitted at \(T_{\rm wait}=0.7\) 
for $1.0\times10^5 \tau_0$ when $\chi_{\rm wait}''$ is measured.
The waiting time is comparable to the sweeping time from $T_{\rm max}$ to 
$T_{\rm min}$. The temperature \( T_{\rm wait} \) 
is set so as to satisfy 
\begin{equation}
T_{\rm c}(1)<T_{\rm wait}<T_{\rm c}(2).
\end{equation}

In Fig.~2 there is obvious hysteresis, 
which does not disappear even in the case of 
slower rates (to $2.0\times10^{-6} T/\tau_0$) and 
longer periods (to $10^3 \tau_0$). 
An important result of the simulation is {\it reinitialization}, 
{\it i.e.}, \( \chi_{\rm wait}'' \) merges with \( \chi_{\rm ref}'' \)
in the early stage of the resumed cooling.

In order to visualize the effect of the aging at \( T_{\rm wait} \) more 
clearly, the difference between \( \chi_{\rm ref}'' \) and 
\( \chi_{\rm wait}'' \) is shown in Fig.~3. If we regard the difference 
as a measure of the aging effect, we notice that 
the system behaves as if it quickly forgets the aging in the cooling stage 
({\it reinitialization}) and again remembers when the system is heated back 
near \( T_{\rm wait} \) ({\it memory} effect), 
which implies that dynamics of the GREM is {\it temperature specific}. 

To reveal the mechanism of these effects, \( \chi_0'' \) and \( \chi_1'' \) 
evaluated from \( {\cal M}_0 \) and \( {\cal M}_1 \) 
(\( \chi''=\chi_0''+\chi_1'' \)) are plotted in Fig.~4. 
The rapid increase of \( \chi'' \) just 
after the resumed cooling is brought by \( \chi_0'' \). As for the relaxation 
at \( T_{\rm wait} \), the first layer is quickly equilibrated and 
\( \chi_0'' \) almost retains a constant value because 
$T_{\rm wait}>T_{\rm c}(1)$, while \( \chi_1'' \) decreases since 
$T_{\rm wait}<T_{\rm c}(2)$. In this sense, \( \chi_1'' \) 
is the affected part and \( \chi_0'' \) is the unaffected one. 
As the resumed cooling goes on, the affected part decreases and the 
unaffected part increases because the peak of \( \chi_0'' \) and that of 
\( \chi_1'' \) are near \( T_{\rm c}(1) \) and \(T_{\rm c}(2)\) respectively. 
As the result, \( \chi_{\rm wait}'' \) merges 
with  \( \chi_{\rm ref}'' \) as if the system forgets the aging at 
\( T_{\rm wait} \). But \( \chi_1'' \) again recovers a large 
contribution to \( \chi'' \) and the system remembers the aging 
when the system is reheated to \( T_{\rm wait} \). 

\section{Discussion}
As Hammann {\it et al} have pointed out\cite{comparative}, 
dynamics in glassy materials seems to be classified into 
two distinct types. In the first type, {\it memory } effect and 
{\it reinitialization}, which make the dynamics {\it temperature specific}, 
are observed. Spin glasses belong to this type as mentioned 
in \S\ref{sec:introduction}. In the other type, {\it memory } effect 
exists but {\it reinitialization} does not exist. Since the system is never 
reinitialized after quenching from above \( T_{\rm c} \), the dynamics is 
{\em cumulative}. In this dynamics, the time evolution in the vicinity of 
 \( T_{\rm c} \) is very important. Polymers\cite{PG,PG2}, 
supercooled glycerol\cite{SCG} and the orientational glasses\cite{OG,OG2} 
belong to this type. 

Now let us discuss how these two distinct types of dynamics observed in glassy 
materials are interpreted within the GREM. Time evolution of energy 
distribution \( P_n(E_n,t) \), which is defined as the probability density
that the system is found at time $t$ in one of the states whose energy of the $n$-th layer is $E_n$, is very different according as \( T>T_{\rm c}(n) \) or 
\( T<T_{\rm c}(n) \)\cite{condmat}. In both cases, \( P_n(E_n,t) \) tries to 
approach the equilibrium distribution proportional to 
\( \exp(E_n/T)\rho_n(E_n)=\exp(\{\frac{1}{T}-\frac{1}{T_{\rm c}(n)}\}E_n) \)
with increasing time. But the sign of the exponent is different in the two 
cases. Consequently, \( P_n(E_n,t) \) quickly converges to the 
equilibrium distribution 
\( P^{\rm eq}_n(E_n)=\{\frac{1}{T}-\frac{1}{T_{\rm c}(n)}\}
\exp(\{\frac{1}{T}-\frac{1}{T_{\rm c}(n)}\}E_n) \) if \( T>T_{\rm c}(n) \), 
whereas \( P_n(E_n,t) \) has a peak moving to higher energy level with 
increasing time if \( T<T_{\rm c}(n) \) (time evolution of $P_n(E_n,t)$ 
in the case of \( T<T_{\rm c}(n) \) is shown in Fig.~6, which is 
mentioned more closely later on). In this model, the peak position 
indicates the {\em age} of the layer. This means that the age of layers which 
satisfy \( T>T_{\rm c}(n) \) is kept to be $0$ since the peak of 
\( P^{\rm eq}_n(E_n) \) is located at \( E_n=0 \).

Although the aging process in each layer is {\it cumulative}, 
the dynamics of the GREM is {\it temperature specific}. 
Now let us discuss what happens in the case of \( L \gg 1 \). 
For a given temperature \( T< T_{\rm c}(L) \), 
there exists the \( n \)-th layer which satisfies 
\( T_{\rm c}(n-1)<T<T_{\rm c}(n) \). As discussed in ref.~\citen{condmat}, 
the crucial point is the fact that the layers below $n$ are quickly 
equilibrated and do not contribute to slow dynamics, 
while those above $n$ are almost quenched and they behave 
as if the time evolution stops. This means that the \( n \)-th layer is the 
activated one and mainly dominates slow dynamics of the system. 
This activated layer changes with temperature, that causes 
{\it temperature specific} dynamics of the GREM. 
Memory of aging at a given temperature is stored in the corresponding 
activated layer as a peak position of \( P_n(E_n,t) \). 
The memory is preserved while the system is cooled since the layer becomes 
frozen, and is reinitialized if the system is heated up to 
a certain temperature \( T+\Delta T>T_{\rm c}(n) \) because the peak of 
\( P_n(E_n,t) \) created at \( T \) is destroyed. 

In the Single-layer Random Energy Model (GREM with $L=1$), 
the dynamics is {\it cumulative} because changes of the activated 
layer do not occur. In Fig.~5, the relaxation of \( \chi'' \) 
during positive \( T \)-cycle in this model is shown. 
The relaxation highly proceeds while the system is heated up, 
which is usually observed in {\it cumulative} systems\cite{PG2}. 
On the other hand, relaxation of \( \chi'' \) is {\it reinitialized} 
by positive \( T \)-cycle in the GREM\cite{condmat}. 

This understanding of aging phenomena observed in glassy materials is 
just the same as that obtained from studies of mean-field 
spin glasses\cite{Cugliandolo}, in which it has been concluded that dynamics 
of mean-field models with a one-step replica symmetry breaking solution
is {\it cumulative} and that with a full replica symmetry breaking solution
is {\it temperature specific}. 

In order to see how activated layers and frozen ones age, 
an example of time evolution of \( P_n(E_n,t) \) is shown in Fig.~6. 
After quenching from an infinitely high temperature, 
the system is kept at \( T=0.45 \). The number of 
layers is three and the transition temperatures are set at 
\( T_{\rm c}(1)=0.6 \), \( T_{\rm c}(2)=0.8 \) and \( T_{\rm c}(3)=1.0 \), 
so that the \( 1 \)-st layer is the activated one and 
the \( 2 \)-nd and the \( 3 \)-rd layers are frozen. 
We notice that the shifting speed of peak position of \( P_n(E_n,t) \) 
in the frozen layers is slower than that in the activated layer. 
In fact, it can be proved analytically that 
in the case of \(t\gg 1\) and \( T_{\rm c}(n-1)<T<T_{\rm c}(n) \), 
the energy distribution \( P_n(E_n,t) \) satisfies the scaling, 
\begin{equation}
P_k(E_k,t)={\overline P}_k(E_k-R_k T \log t)\hspace{1cm}(k\ge n),
\end{equation}
\begin{equation} 
R_k \equiv \left\{ 
  \begin{array}{cl}
1&\mbox{($ k=n $)}, \\
\displaystyle{\prod_{l=n}^{k-1} \frac{T}{T_{\rm c}(l)}}
&\mbox{($ k>n $)}.
\end{array}\right.
\end{equation}
The validity is clear from the scaling plots shown in the insets of Fig.~6. 
The scaling implies that the age of the \( k \)-th ($k>n$) frozen layer 
at $t$ is nearly equal to that of the activated layer at \( t^{R_k} \), 
if age of each layer is measured by the peak position of the 
energy distribution. This frozen property becomes very important at 
experimental time scale \( t_{\rm exp} \), which is considered to be 
\( 10^{13}\le t_{\rm exp}\le 10^{18} \) in ordinary spin glasses 
in unit of the microscopic time of the system. 
%if the microscopic time of the system is used as the unit of time. 

\section*{Acknowledgment}
The numerical simulations were made on an Origin 2000 at Division 
of Physics, Graduate school of Science, Hokkaido University.

\newpage
\noindent
{\bf \large FIGURE CAPTIONS}

\vspace*{3mm}\noindent
Fig.~1  Structure of the GREM with \( L=2 \) and 
        \(N=5\). The open circles represent accessible states 
	of the system. 

\vspace*{3mm}\noindent
Fig.~2  Out-of-phase ac-susceptibility \( \chi'' \) measured in the GREM with 
	$L=2$, $T_{\rm c}(1)=0.5$ and $T_{\rm c}(2)=1.0$. The amplitude 
	and the period of the applying ac-field are \( 0.1 \) and $100 \tau_0$,
        respectively. The system is cooled (and reheated) at the rate of 
	$2.0\times10^{-5} T/\tau_0$. In the case of \( \chi_{\rm wait}'' \) 
	measurement, the cooling is intermitted at \( T_{\rm wait}=0.7 \) 
	for \( 1.0\times10^5 \tau_0 \). 
	The solid line, open circles and full circles correspond to 
        $\chi_{\rm ref}''$, $\chi_{\rm wait}''$ (cooling) and 
	$\chi_{\rm wait}''$ (reheating), respectively. 

\vspace*{3mm}\noindent
Fig.~3  Difference between \( \chi_{\rm ref}'' \) and \( \chi_{\rm wait}'' \)
	in the cooling and the reheating stages. 
        The open circles and the full circles correspond to the 
        cooling and the reheating data, respectively. 

\vspace*{3mm}\noindent
Fig.~4  Out-of-phase ac-susceptibility \( \chi_0'' \) and \( \chi_1'' \) 
	evaluated from ${\cal M}_0$ and ${\cal M}_1$. The solid line, 
	open circles and full circles correspond to 
        $\chi_{\rm ref}''$, $\chi_{\rm wait}''$ (cooling) and 
	$\chi_{\rm wait}''$ (reheating), respectively. 

\vspace*{3mm}\noindent
Fig.~5  Effect of positive $T$-cycle 
        in the Single-layer Random Energy model (the GREM with $L=1$) 
	with \( T_{\rm c}=1.0 \). The ac-field with the peak amplitude 
	\( 0.1 \) and the period $100 \tau_0$ is applied 
	for the measurement of $\chi''$. After quenching from an infinitely 
	high temperature, the system is kept at \( T=0.5 \). 
	Then a positive temperature 
        perturbation \( \Delta T =0.35 \) is applied at 
	\( t_1=4\times10^3\tau_0  \) and is switched off at 
	\( t_1+t_2=1.4\times10^4 \tau_0 \). In the inset, 
        \( t_2 \) part of data is omitted and \( t_1 \) and 
        \( t_3 \) parts are connected for comparison with the unperturbed 
        data (dotted line). 

\vspace*{3mm}\noindent
Fig.~6  Time evolution of \( P_n(E_n,t) \) of the GREM with $L=3$, 
        $T_{\rm c}(1)=0.6$, $T_{\rm c}(2)=0.8$ and $T_{\rm c}(3)=1.0$. 
	After quenching from an infinitely high temperature, 
	the system is kept at \( T=0.45 \) and \( P_n(E_n,t) \) is 
	measured at $t=10^{4.5},10^{5.0},10^{5.5},\ldots,10^8 \tau_0$ 
	(from left to right). 
	In the inset, \( P_n(E_n,t) \) is plotted as a function of 
	$E_n-R_n T \log t$, where $R_1=1.0$, $R_2=0.75$ and $R_3=0.42$
	(see text). 

\newpage
\begin{center}
\epsfile{file=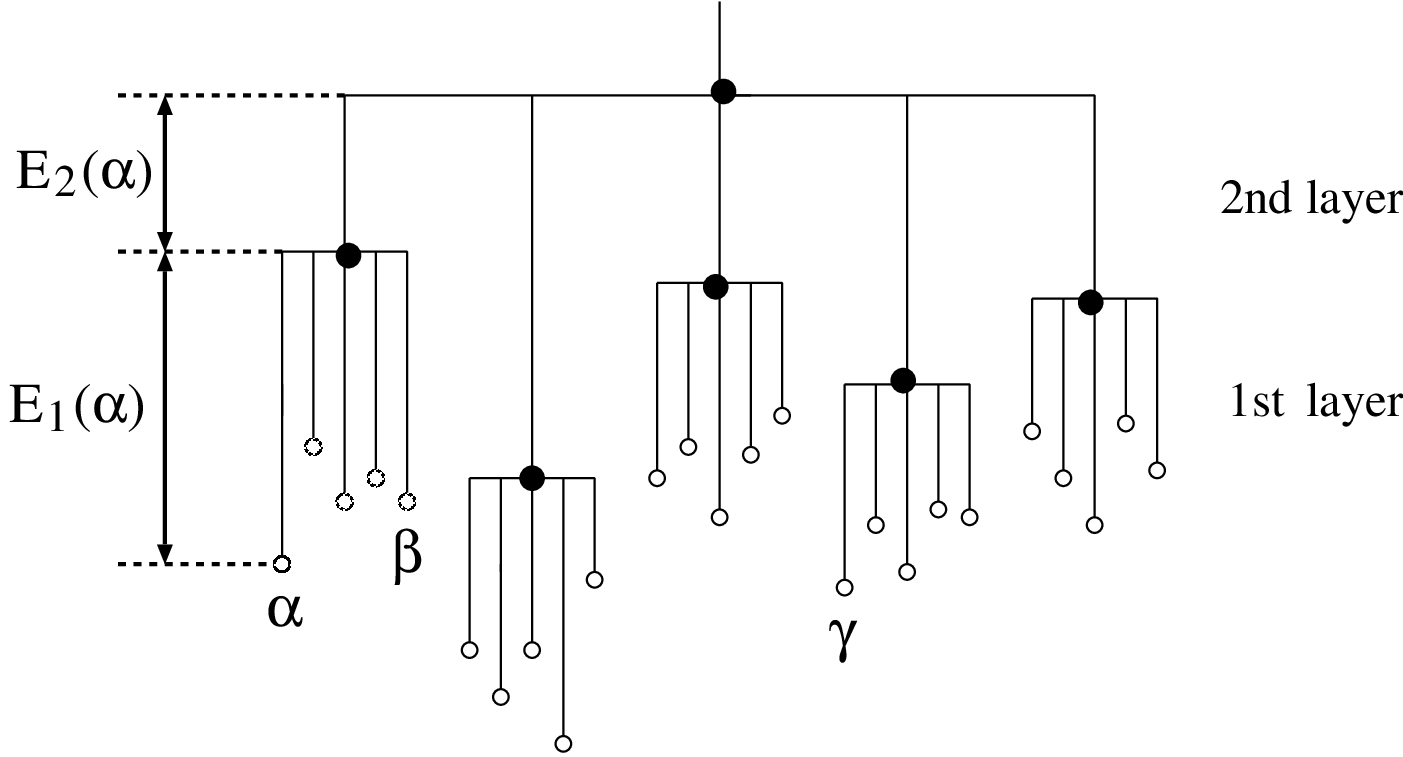,width=15.5cm}
\end{center}
\vspace{3cm}
\begin{center}
{\LARGE Fig.1}
\end{center}
\newpage
\begin{center}
\epsfile{file=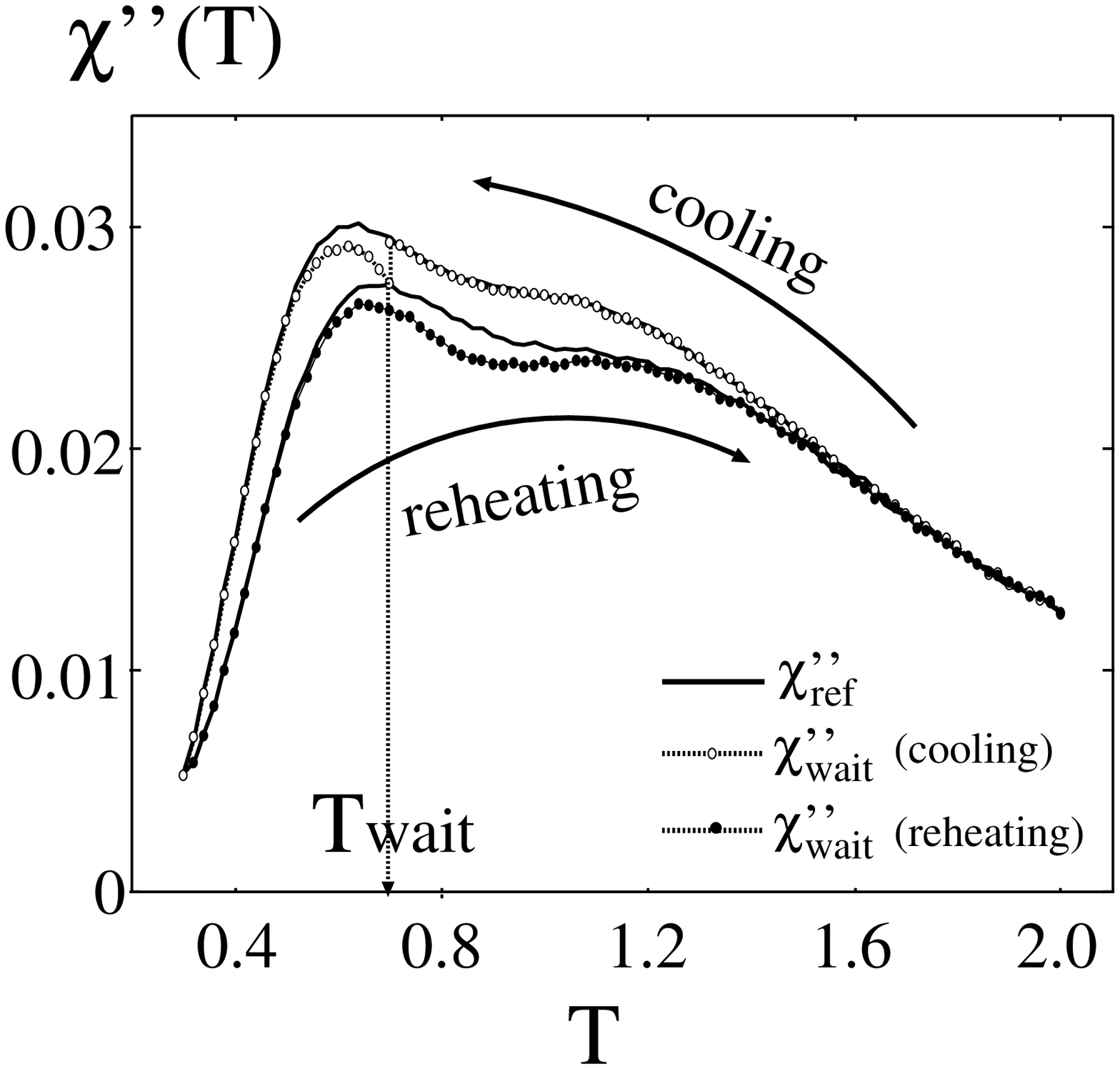,width=15.5cm}
\end{center}
\vspace{3cm}
\begin{center}
{\LARGE Fig.2}
\end{center}
\newpage
\begin{center}
\epsfile{file=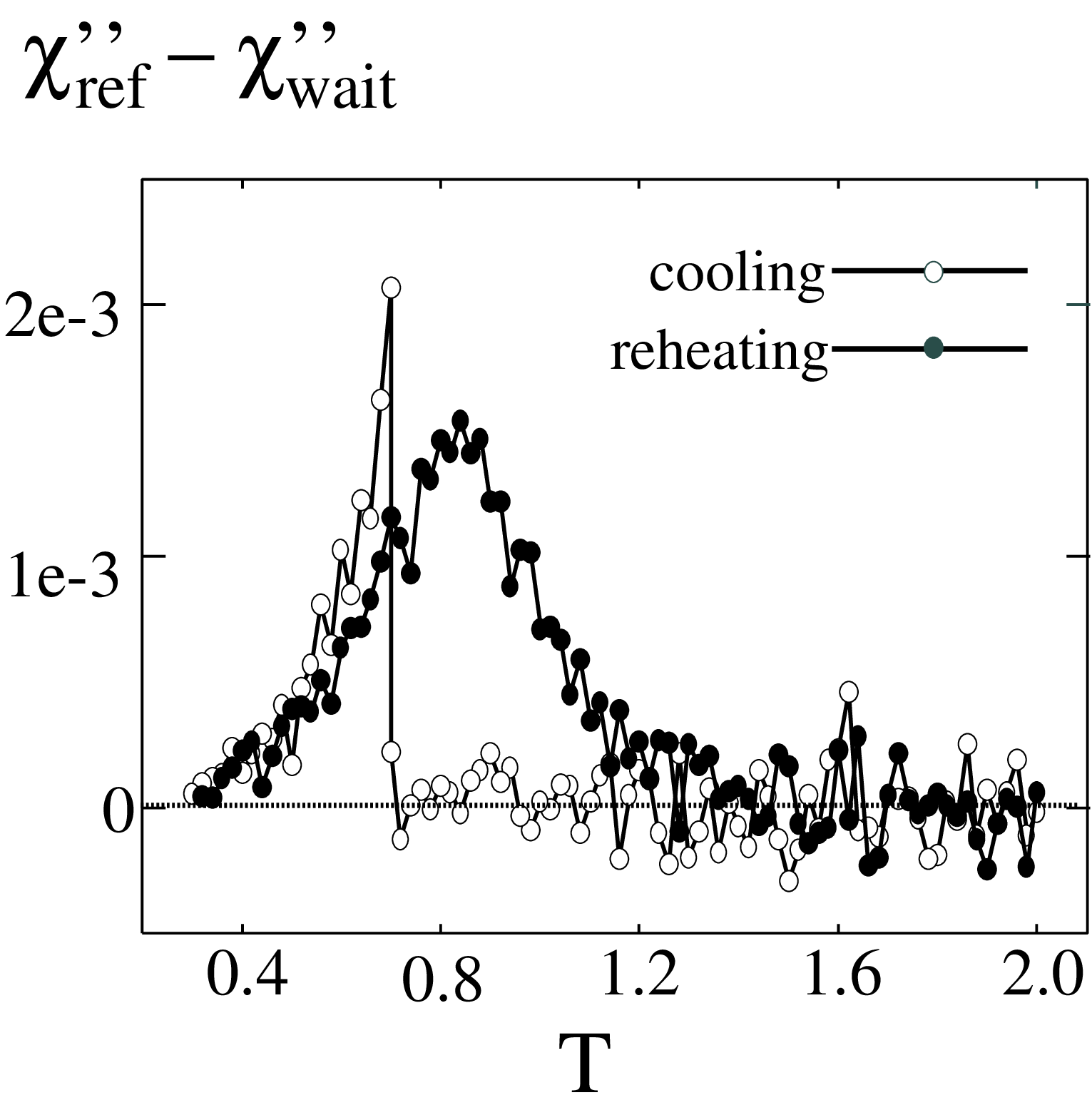,width=15.5cm}
\end{center}
\vspace{3cm}
\begin{center}
{\LARGE Fig.3}
\end{center}
\newpage
\begin{center}
\epsfile{file=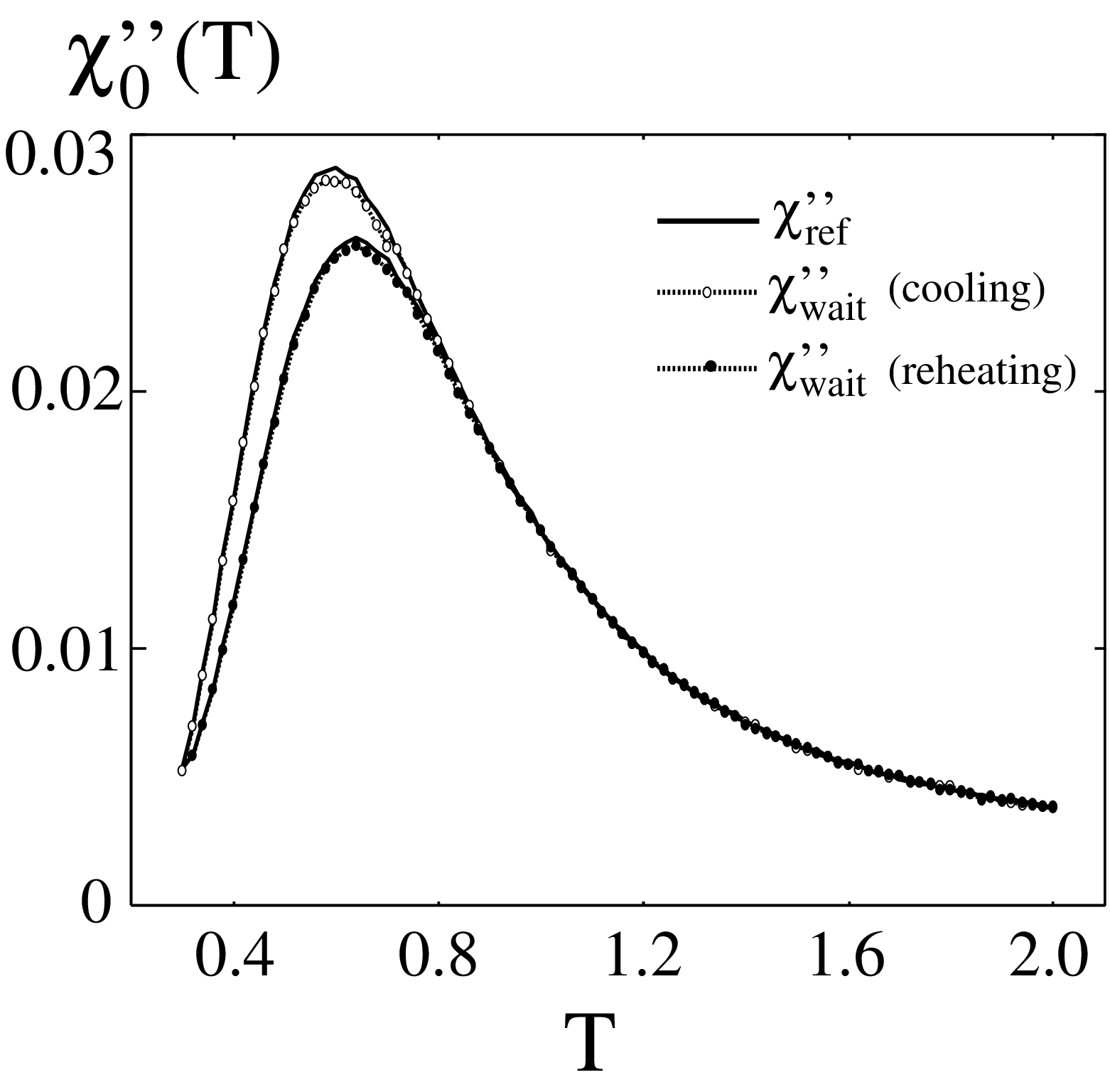,width=15.5cm}
\end{center}
\vspace{3cm}
\begin{center}
{\LARGE Fig.4(a)}
\end{center}
\newpage
\begin{center}
\epsfile{file=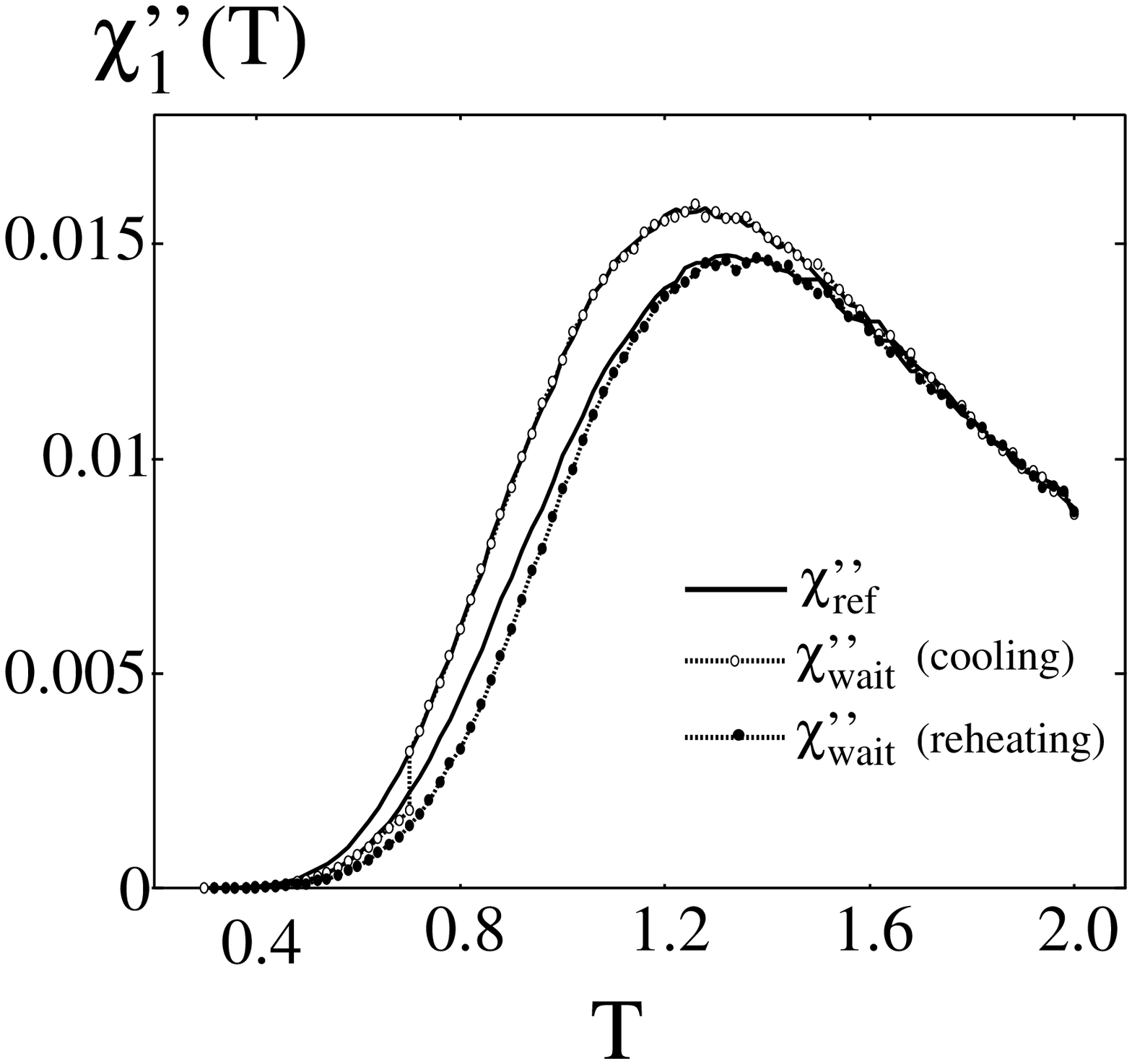,width=15.5cm}
\end{center}
\vspace{3cm}
\begin{center}
{\LARGE Fig.4(b)}
\end{center}
\newpage
\begin{center}
\epsfile{file=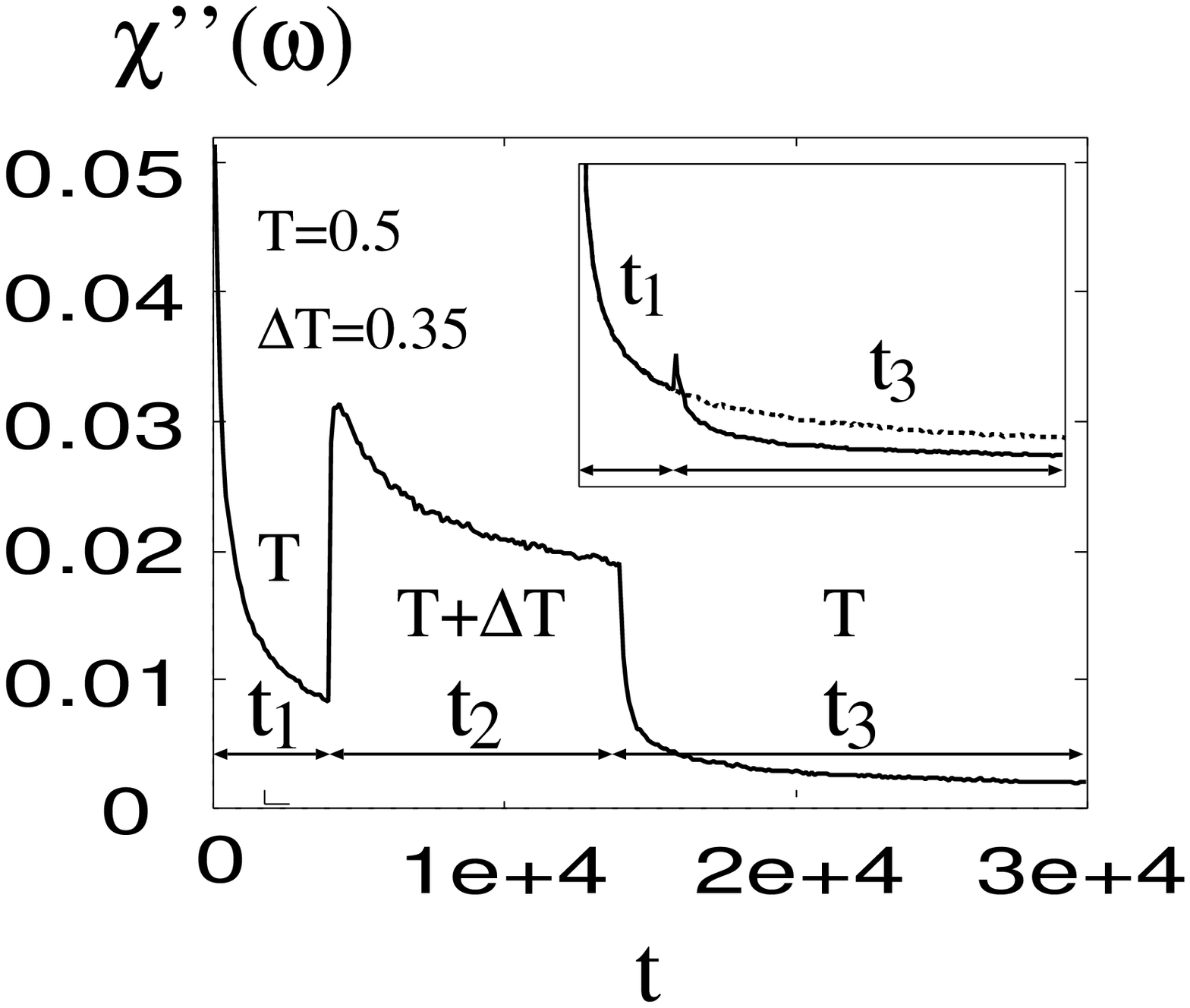,width=15.5cm}
\end{center}
\vspace{3cm}
\begin{center}
{\LARGE Fig.5}
\end{center}
\newpage
\begin{center}
\epsfile{file=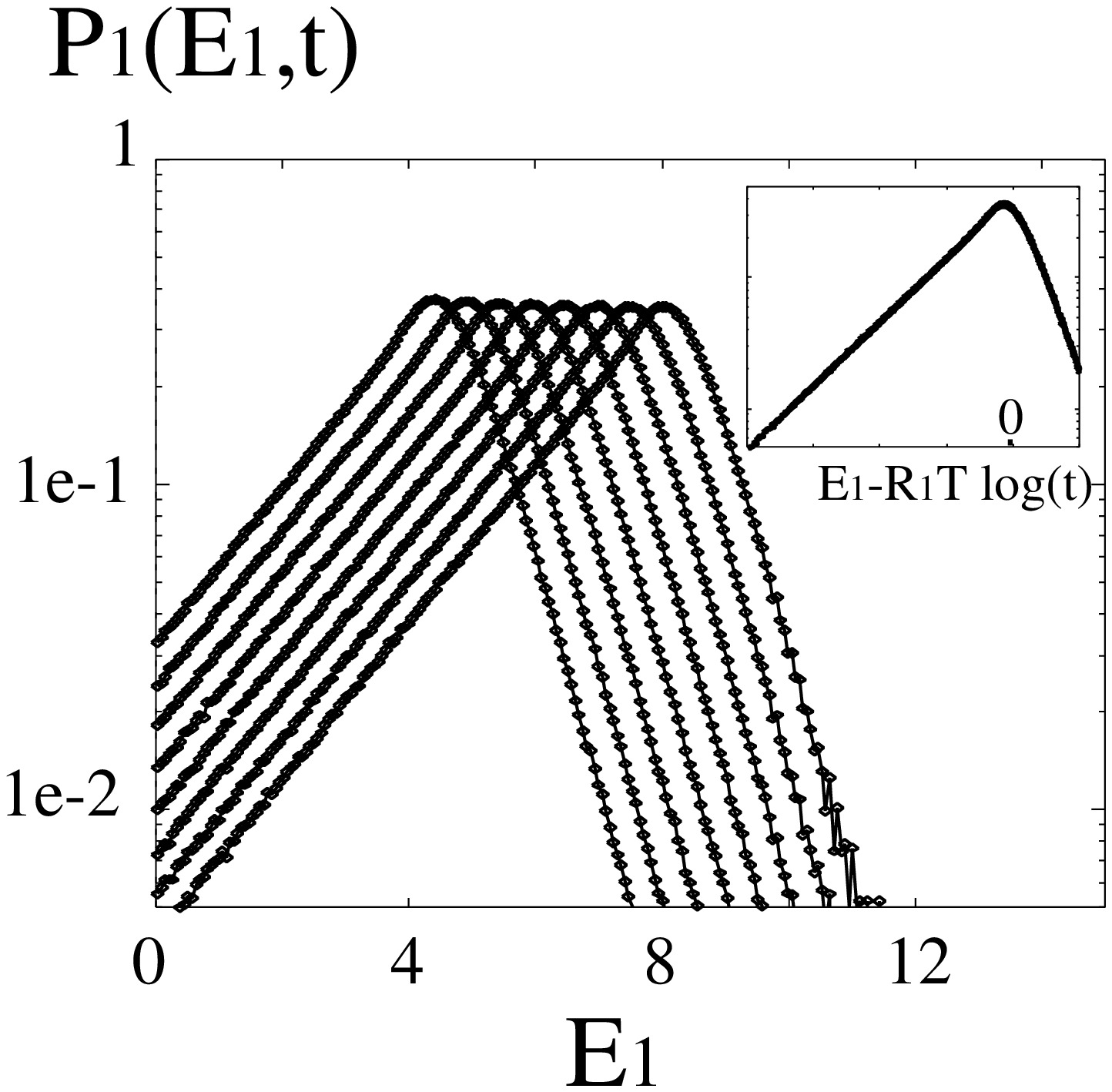,width=15.5cm}
\end{center}
\vspace{3cm}
\begin{center}
{\LARGE Fig.6(a)}
\end{center}
\newpage
\begin{center}
\epsfile{file=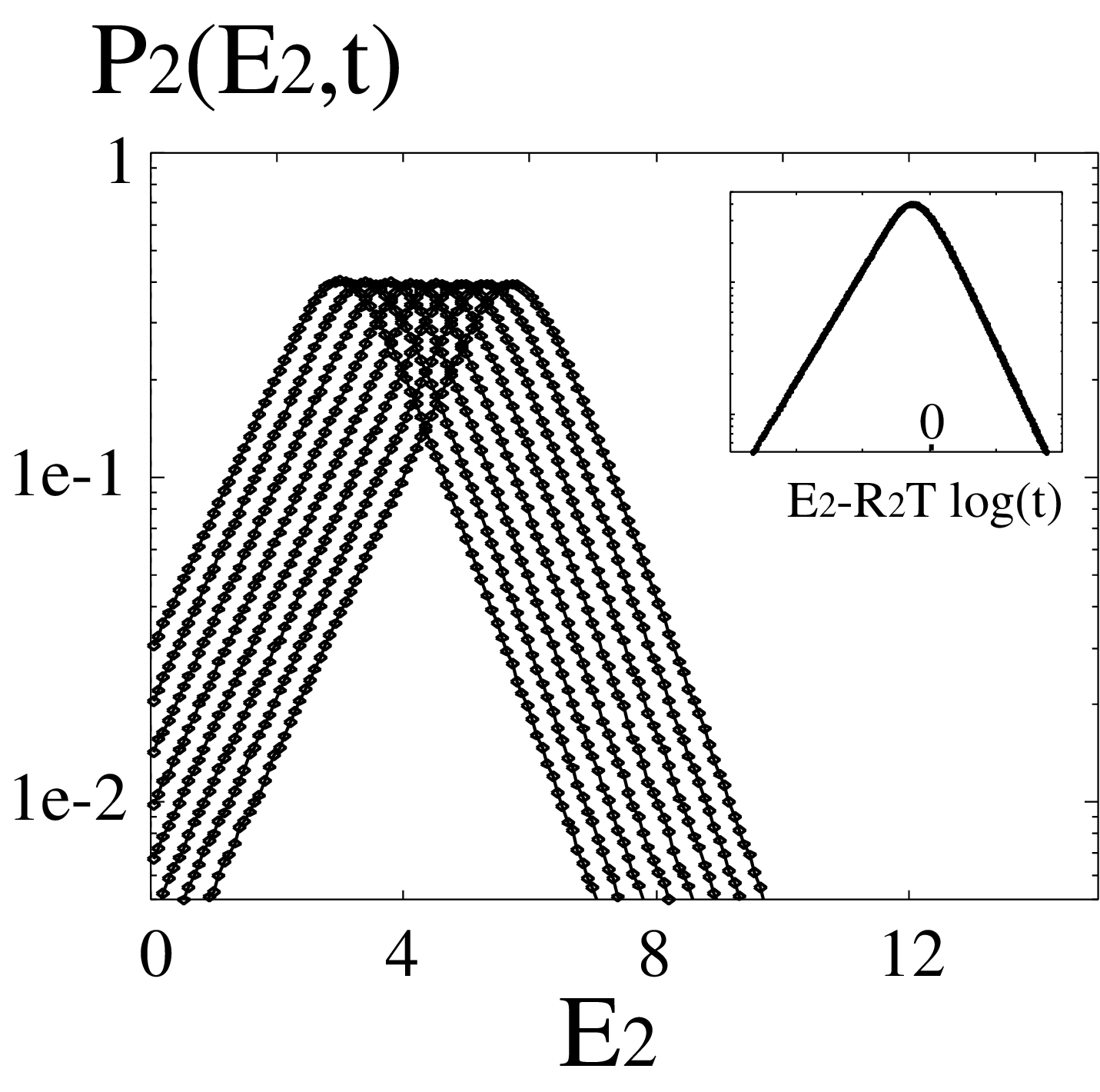,width=15.5cm}
\end{center}
\vspace{3cm}
\begin{center}
{\LARGE Fig.6(b)}
\end{center}
\newpage
\begin{center}
\epsfile{file=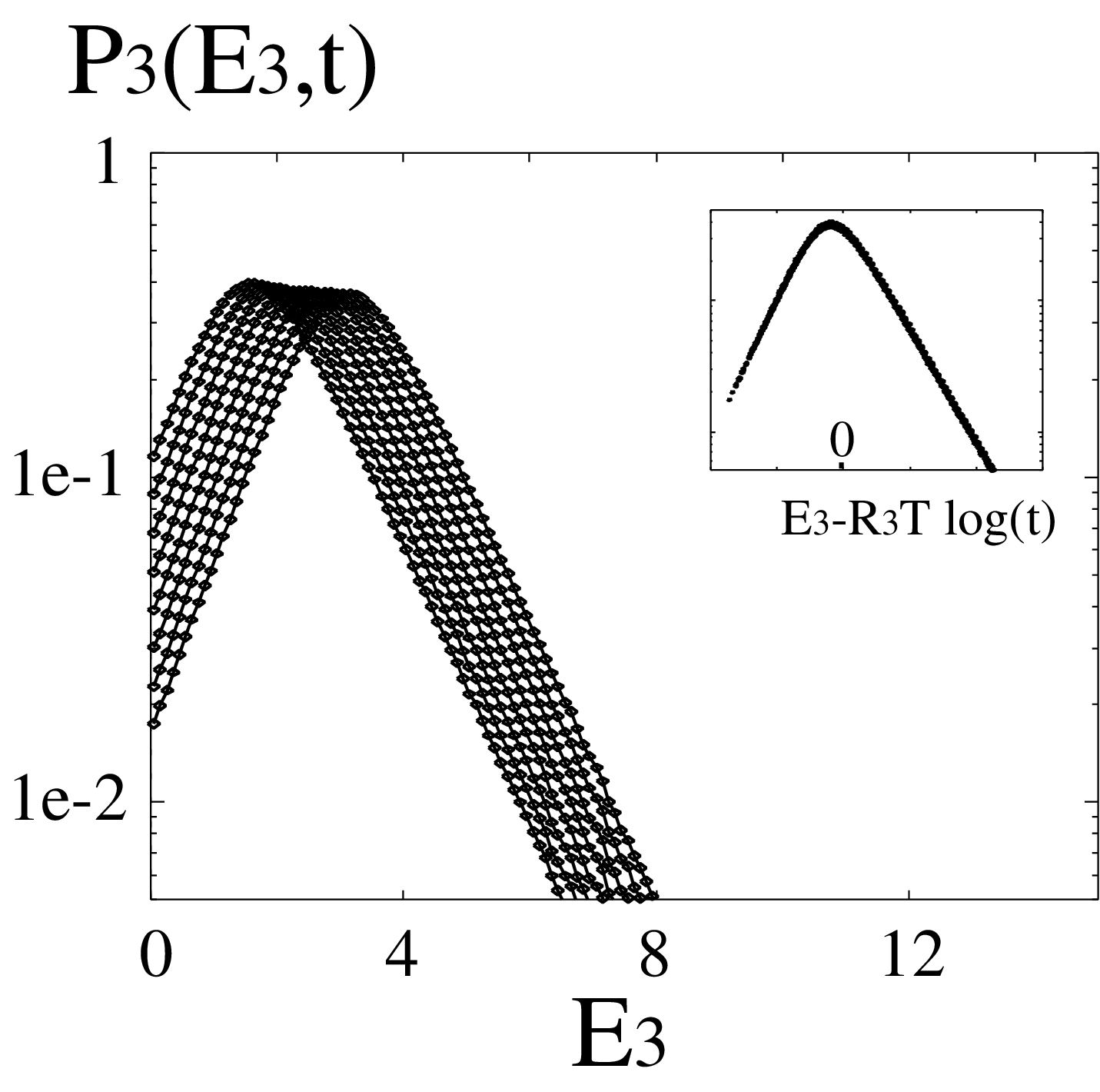,width=15.5cm}
\end{center}
\vspace{3cm}
\begin{center}
{\LARGE Fig.6(c)}
\end{center}

\end{document}